# Highly efficient intracellular chromobody delivery by mesoporous silica nanoparticles for antigen targeting and visualization in real time

Hsin-Yi Chiu[a], Wen Deng[b], Hanna Engelke[a], Jonas Helma[b], Heinrich Leonhardt[b,]*, and Thomas Bein[a,]*

[a] Department of Chemistry and Center for NanoScience (CeNS), University of Munich (LMU), Butenandtstrasse 5-13 (E), 81377 Munich, Germany

[b] Department of Biology II and Center for NanoScience (CeNS), University of Munich (LMU), Grosshadernerstrasse 2, 82152 Planegg-Martinsried, Germany

* Address correspondence to: bein@lmu.de, h.leonhardt@lmu.de

## Abstract

Chromobodies have recently drawn great attention as bioimaging nanotools. They offer antigen binding specificity and affinity comparable to conventional antibodies, but much smaller size and higher stability. Importantly, chromobodies can be used in live cell imaging for highly specific spatio-temporal visualization of cellular processes. To date, functional application of chromobodies requires lengthy genetic manipulation of the target cell. Here, we developed multifunctional large-pore mesoporous silica nanoparticles (MSNs) as nanocarriers to *directly* transport chromobodies into living cells for antigen-visualization in real time. The multifunctional large-pore MSNs feature high loading capacity for chromobodies (70 μg chromobody/mg MSNs), and are efficiently taken up by cells (after 2 h of incubation with chromobody-loaded MSNs (MSN-Ca-Cbs) more than 90 % of the cells have taken up the MSN-Ca-Cbs). By functionalizing the internal MSN surface with nitrilotriacetic acid-metal ion (NTA-$M^{2+}$) complexes, we could control the release of $His_6$-tagged chromobodies from MSNs in acidified endosomes. When chromobodies escape from the endosomes through the proton sponge effect generated by their built-in $His_6$-tags, co-localization of signals from fluorescent endogenous antigen and organic dye-labeled chromobodies can be detected. The successful chromobody delivery signals can be observed even at very low effective concentrations of chromobodies introduced to the incubation solution. Various endosomal release triggers (fusogenic peptide INF7, acid shock, DMSO, chloroquine) were further examined to enhance the endosomal release of chromobodies. The results showed that short incubation with DMSO or chloroquine with MSN-Ca-Cb pretreated cells at room temperature results in a 12 – 18 fold increase of chromobody release efficiency. Hence, by combining the two nanotools, chromobodies and MSNs, we established a new powerful approach for chromobody applications in living cells.

**Introduction**

Today, antibodies are considered to be the most powerful tools for specific visualization of cellular compartments at the molecular level aimed at the study of cellular processes. They are indispensable for proteomic analyses, protein localization and detection of posttranslational modifications. However, the application of full-length antibodies is restricted to fixed, meaning dead cells, since the massive sizes (~ 150 kD) and complex folding structures including intermolecular disulphide bridges limit their use in living cells *via* the transient expression approach or *via* direct delivery. As a result, the idea of engineering recombinant small antibodies for real time dynamic protein tracing in living cells has received much attention. A variety of recombinant small antibodies including immunoglobulin (Ig) derived Fab (~ 50 kD) and scFv (~ 25 kD), as well as non-Ig derived monobody (~ 10 kD) and affibody (~ 6.5 kD) protein scaffolds have been generated in the last decades for this purpose[1]. Nanobodies (~ 14 kD) are the single-domain antigen-binding fragments derived from camelid's single-chain IgG[2]. They have a binding affinity and specificity similar to conventional antibodies, but are much smaller in size and exhibit higher stability. When conjugated with fluorescent proteins or organic dyes, the fluorescent nanobodies, named chromobodies, become molecular probes that can trace the dynamics of endogenous cellular structures in living cells. Chromobodies have successfully shown their antigen detection efficacy on cytoskeleton, histone protein and DNA replication complexes, and have revealed the spatio-temporal protein changes during cell cycles[3]. In our previous report[4], HIV-specific chromobodies have been generated and used for real time visualization of HIV assembly in living cells. These studies demonstrate that chromobodies are promising protein reporters for the study of cellular processes in living cells. However, to date the application of chromobodies for live cell imaging was limited due to the need to introduce them genetically, followed by subsequent cytosolic expression. To broaden the flexibility and use of chromobodies in biomedical applications (e.g., manipulation of cell function for disease treatment), direct intracellular delivery of the molecular probes would be highly desirable. However, intracellular protein delivery is challenging firstly because the large size of proteins leads to difficulties with passive diffusion through the cell membrane or with endocytosis. The following endosomal trapping of internalized proteins further limits the protein functions in cells. A few studies of non-carrier intracellular protein delivery aimed to enhance the cellular uptake efficiency in combination with endosomolytic agents and therefore to increase the protein delivery efficacy[5, 6]. For example, Erazo-Oliveras *et al.* co-incubated dimerized cell-penetrating peptide TAT (dfTAT) with proteins (EGFP and Cre recombinase, etc.) in culture media, the results show that proteins were successfully crossing the cell membrane with the help of dfTAT and presenting their functions in cytosols[5]; D'Astolfo *et al.* used hyperosmolality NaCl buffer to induce the high efficient micropinocytosis of proteins. In combination with a transduction compound (a propanebetaine), intracellcular release was achieved[6]. Alternatively, using carriers such as cationic polyplexes with high cellular uptake efficiency for proteins may improve the generally low efficiency of protein uptake. However, the complex structures and wide variety of surface charges of proteins make it difficult to design a general carrier for universal protein delivery[7]. The second challenge of protein delivery is that proteins are susceptible. Proteins must maintain their tertiary structure to preserve their functionality during the delivery process.

To address these bottlenecks, we surmised that mesoporous silica nanoparticles (MSNs) could be promising candidates for serving as efficient and versatile protein delivery vehicles. For example, their surface can be modified with different functionalities and charges to accommodate different proteins; their pore sizes are tunable to fit different cargo sizes; their framework is stable and can effectively protect cargos from environmental degradation, and they can be efficiently taken up by cells. In the past years, the development of MSNs for biomedical applications has greatly increased. The achievements include delivery of chemotherapeutic agents for cancer therapy [8], intracellular protein delivery for manipulation of cell function[9, 10], and oligonucleotide delivery for gene therapy[11, 12]. These studies demonstrate that MSNs can efficiently control the release of cargos in the target tissue/cells as well as effectively protect cargos from degradation. Importantly, MSNs were found to be biocompatible within certain concentration ranges[13].

In this study, we synthesized multifunctional large-pore MSNs for intracellular chromobody delivery. Metal chelate complexes were covalently attached on the internal silica surface and used for pH-responsive coordination binding of $His_6$-tagged chromobodies. The binding affinity and pH-stimulated release of various metal ions ($Fe^{2+}$, $Co^{2+}$, $Ni^{2+}$, $Cu^{2+}$, $Zn^{2+}$ and $Ca^{2+}$) with $His_6$-tagged chromobodies were examined by colorimetric measurements *in vitro*. Mouse embryonic fibroblasts (MEFs) expressing EGFP fused to LMNA (an inner nuclear membrane protein) (MEF-G-LMNA) were generated for the detection of GFP-specific chromobody release and function in intracellular delivery experiments. The successful chromobody delivery, release from the endosomes and binding to the target structures could be confirmed by the fluorescence co-localization signals of EGFP and chromobodies on the LMNA structure.

## Results and Discussion

**Synthesis and characterization of large-pore MSN**

Mercapto-functionalized MSNs (MSN-SH) were synthesized using a sol-gel templating approach in a neutral pH reactive solution according to a modified protocol from the literature[14]. During the synthesis process (Figure 1a) the structure directing agent cetyltrimethylammonium ($CTA^+$) formed micelles in the initial solution, silica precursors TEOS and MPTES were then adsorbed and co-condensed around the micelles because of the electrostatic attraction between $CTA^+$ and negatively charged silica. In this neutral pH synthesis solution, the counter ions $Tos^-$ compete with the silicate oligomers for association with the positively charged micelles, which results in sparse silica condensation and consequently the large stellate pore structure formation (Figure 1b)[14]. Mercapto-functional groups were introduced homogeneously into the silica framework *via* cocondensation for the purpose of further functionalization. According to the $N_2$ sorption analysis (Figure 1c), MSN-SH has a fairly wide pore size distribution from 10 nm to 20 nm, a BET surface area of 670 $m^2$/g and a large pore volume of 3.06 $cm^3$/g. With these pore dimensions, chromobodies featuring a size of 2 nm x 4 nm[15] are expected to be efficiently loaded into the mesopores. The hydrodynamic particle size (Figure 1d) measured by DLS was 100 – 200 nm. This particle size range is considered to be favorable for endocytosis[16].

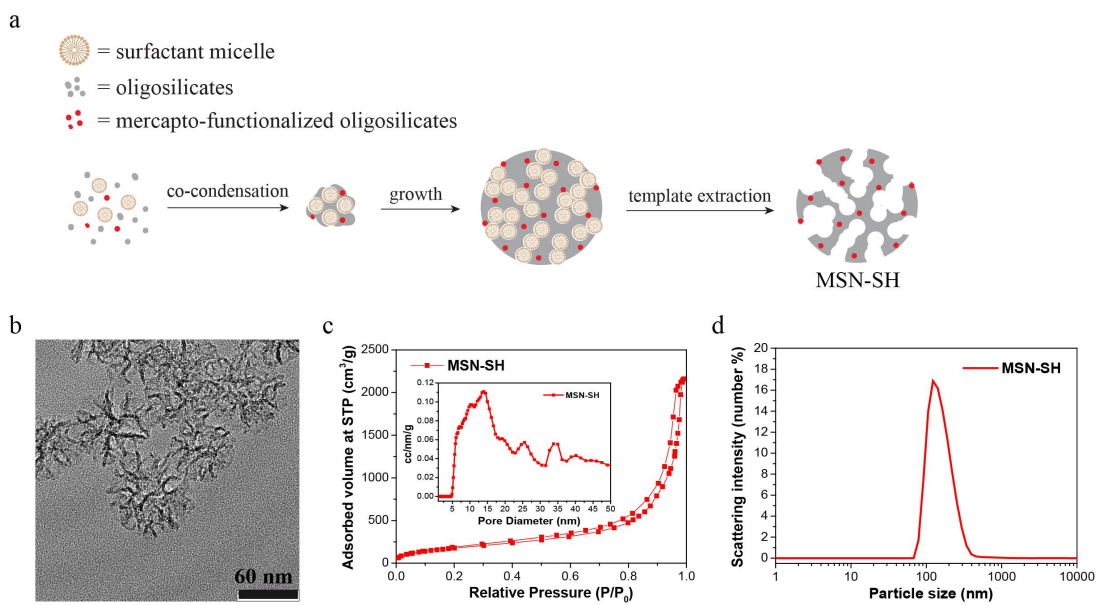

**Figure 1. Synthesis and characterization of MSN-SH.** (a) MSN-SH was synthesized through a modified protocol described earlier in the literature[14]. MSN-SH was synthesized by co-condensation of oligosilicates (from TEOS) and mercapto-functionalized oligosilicates (from MPTES) in a neutral pH reaction mixture. (b) TEM image of MSN-SH. The high contrast areas indicate the dense silica backbones whereas the low contrast areas indicate the pore structure. The average particle size is about 100 nm according to the TEM image. (c) Nitrogen sorption isotherm (outer figure) of MSN-SH and its corresponding pore size distribution (inner figure) calculated by the NLDFT mode based on the adsorption branch of $N_2$ on silica. (d) Dynamic light scattering (DLS) of hydrodynamic particle size of MSN-SH in EtOH. The average hydrodynamic particle size is around 160 nm.

## NTA-metal ion complex modified MSNs for controlled uptake and release of chromobodies

To control chromobody loading and release from the silica framework, nitrilotriacetic acid -metal ion complexes (NTA-$M^{2+}$) were attached to the MSN surface as pH-responsive linkers. The recombinant GFP-specific chromobody possesses a built-in $His_6$-tag on its C-terminus for purification purposes. This $His_6$-tag can be used as a tether to conjugate chromobodies onto NTA-$M^{2+}$ complexes because there are three potential coordination sites on the histidine molecule: the carboxyl group ($pK_a$ = 1.9), the imidazole nitrogen ($pK_a$ = 6.1) and the amino nitrogen ($pK_a$ = 9.1). Among these coordination sites, the imidazole nitrogen is considered to be the primary site for the conjugation with metal ions. The conjugation can be separated by high imidazole concentration buffer elution (free imidazole substitutes the coordination binding of histidine on the NTA-$M^{2+}$ complexes). Alternatively, low pH buffer can also be used for elution since the imidazole nitrogen is protonated under acidic conditions (pH < 6) and therefore acidification results in the detachment of His$_6$-tagged chromobodies from NTA-M$^{2+}$ complexes.

To carry out the MSN modification, MSN-SH was converted to carboxyl-functionalized MSN (MSN-COOH) by reacting MSN-SH with 6-maleimidohexanoic acid in ethanol. The MSN-COOH was then conjugated with Nα,Nα-bis(carboxymethyl)-L-lysine hydrate (NTA-lysine) via an EDC-sulfo NHS coupling approach (Figure 2a). The 1700 cm$^{-1}$ peak attributed to a C=O vibration in the IR spectrum (Figure 2b, red line) indicated the successful coupling of 6-maleimidohexanoic acid with MSN-SH, while the conjugation of MSN-COOH and NTA-lysine was verified by the increased secondary amide bond signal at 1650 cm$^{-1}$ in the IR spectrum (Figure 2b, blue line). After this two-step modification, the final NTA-conjugated MSNs (MSN-NTA) still exhibit the desired large-pore structure as well as colloidal stability (Figure S1). To compare the effect of different NTA-M$^{2+}$ complexes for His$_6$-tagged chromobody conjugation, various metal ions: Fe$^{2+}$, Co$^{2+}$, Ni$^{2+}$, Cu$^{2+}$, Zn$^{2+}$ and Ca$^{2+}$ were immobilized onto MSN-NTA, yielding NTA-M$^{2+}$-complex-modified MSNs (MSN-M$^{2+}$) (Figure 2c). Chromobodies were then loaded on the MSN-M$^{2+}$ carriers (MSN-Fe$^{2+}$, MSN-Co$^{2+}$, MSN-Ni$^{2+}$, MSN-Cu$^{2+}$, MSN-Zn$^{2+}$ and MSN-Ca$^{2+}$) as well as MSN-SH (as control group) in 0.05 M Tris-acetate buffer (pH 8). Subsequently, chromobody loading and release tests were performed via measurements of the fluorescence intensity (emission at 669 nm) of the loading/release supernatants. The fluorescent intensity of a dilution series of pure chromobodies was measured as standard curve for the following quantification. Based on the results shown in Figure 2d, all samples (MSN-Fe$^{2+}$, MSN-Co$^{2+}$, MSN-Ni$^{2+}$, MSN-Cu$^{2+}$, MSN-Zn$^{2+}$, MSN-Ca$^{2+}$ and MSN-SH; for the latter, see discussion below) exhibited a similarly high chromobody loading capacity of approximately 70 μg/mg MSN. This loading capacity corresponds to about 600 chromobody molecules per MSN. After incubation of the chromobody-loaded MSNs (MSN-M-Cb) in pH 7 buffer for 16 h, all the samples showed on average only 5 % release of the loaded chromobodies (Figure 2e). However, performing the chromobody release in pH 5 buffer, MSN-Ni$^{2+}$, MSN-Cu$^{2+}$, MSN-Zn$^{2+}$ and MSN-Ca$^{2+}$ exhibited significantly higher chromobody release efficiency than in pH 7 (Figure 2e). In general, the extent of release of His$_6$-tagged proteins from the NTA-M$^{2+}$ moieties is determined by two factors: (i) the stability of the metal chelation by NTA, and (ii) the stability of the coordination between metal ions and histidine molecules. The general stability order of NTA-metal ion and histidine-metal ion complexes is: Cu$^{2+}$ > Ni$^{2+}$ > Zn$^{2+}$ > Co$^{2+}$ > Fe$^{2+}$ > Ca$^{2+}$[17, 18]. The results shown here indicate that Cu$^{2+}$, Ni$^{2+}$, Zn$^{2+}$ and Ca$^{2+}$ are able to allow for the chromobody release in an acidic environment at pH 5, while $Cu^{2+}$, $Ni^{2+}$ and $Zn^{2+}$ show a rather high stability when incorporating to the MSN-NTA system. By contrast, $Co^{2+}$ and $Fe^{2+}$ showed no significant differences of chromobody controlled release in pH 7 and pH 5 buffers due to their weak complexing stability with either NTA or histidine. Interestingly, although the coordination stability of NTA-$Ca^{2+}$ is low, $His_6$-tagged chromobodies still bind sufficiently to MSN-$Ca^{2+}$ in pH 7 buffer. Practically, MSN-$Ca^{2+}$ is favored for chromobody delivery because the decomposition of the NTA-Ca-chromobody complex starts already at pH 6 (data not shown), therefore more chromobodies can be released in the early endosome stage. Moreover, small amounts of $Ca^{2+}$ are considered to be biocompatible and non-toxic.

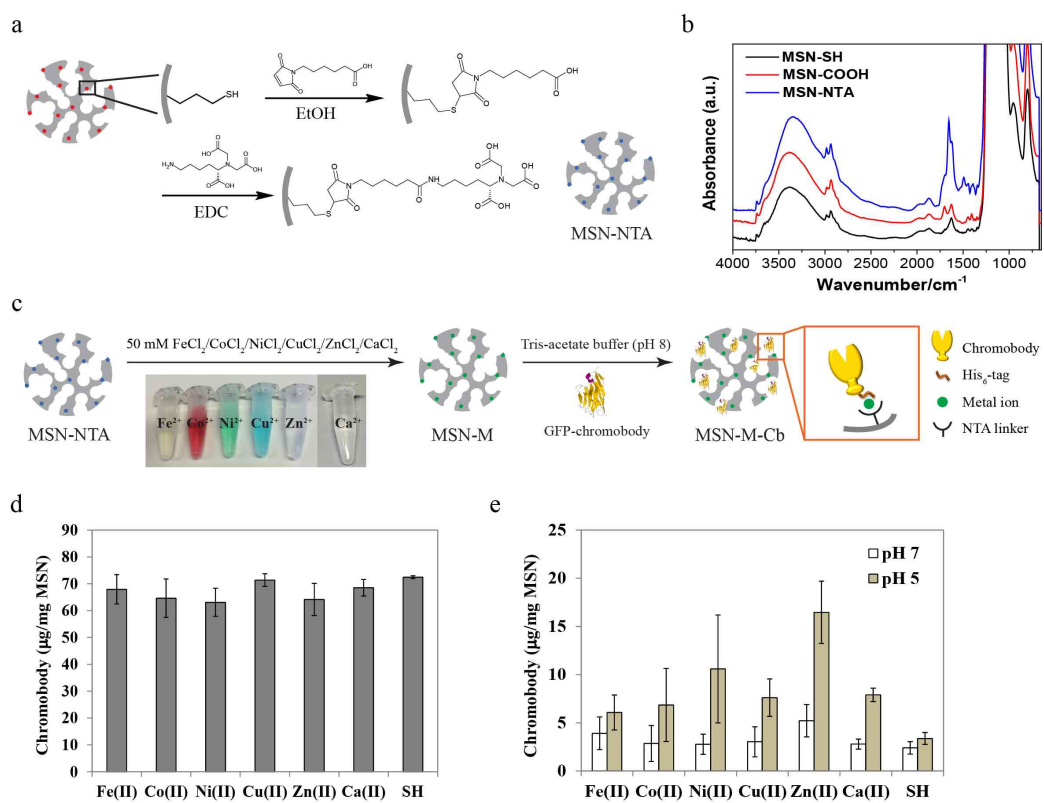

**Figure 2. Surface modification of MSN and chromobody (atto 647N labeled) controlled release test *in vitro*.** (a) Modification process of MSN-SH to yield MSN-NTA. (b) IR spetra of the functionalized MSNs during each modification step. MSN-SH (black), MSN-COOH (red) and MSN-NTA (blue). (c) MSN-NTA particles were treated with different metal ion solutions (50 mM), respectively, at room temperature for 6 h. The free metal ions were then washed away with water, and the resulting metal-immobilized MSNs (MSN-$M^{2+}$) were incubated with chromobody loading buffers (100 μg chromobody/ml in Tris-acetate buffer, pH 8) at 4 °C for 2 h to bind the chromobodies to MSNs. (d) Chromobody loading capacity of different metal-immobilized MSNs. MSN-SH served as control group in the *in vitro* controlled release test. (e) Chromobody loaded MSN particles (MSN-M-Cb) were dispersed in PBS buffer (with pH 7 or pH 5) at 37 °C for 16 h. Afterwards, the supernatants were collected and the amounts of chromobody released were measured using colorimetric analysis. Pure chromobodies were diluted in PBS buffer at different concentrations for creating the standard curve. Error bar: standard error.

The control group MSN-SH particles showed high chromobody loading capacity and low chromobody release in both pH 7 and pH 5 even without any capping system or chemical conjugation. This indicates that the negatively charged silica framework (zeta potential: -13.7 mV at pH 5, -30.8 mV at pH 7) exhibits non-specific binding to chromobodies. The MSN-$M^{2+}$ carriers we developed here are not only useful for His$_6$-tagged chromobody controlled release, but also promising for universal His-tagged protein delivery to living cells. Several groups have previously shown the successful binding and release of His-tagged proteins on NTA-$Ni^{2+}$ complex modified silica substrates either for protein delivery or for biosensor detection[9, 10, 19]. Similarly, these proteins could be attached to the large internal surface of our MSN-$M^{2+}$ carrier system.

MSN-$Ni^{2+}$, MSN-$Zn^{2+}$ and MSN-$Ca^{2+}$ (primary carrier) were used for the following intracellular chromobody delivery study in MEF-G-LMNA cells. To address possible concerns of cytotoxicity of the dissociated metal ions after uptake of the particles into the cells, we performed MTT assays after incubation of MSN-$Ni^{2+}$, MSN-$Zn^{2+}$ and MSN-$Ca^{2+}$ particles with the MEF cells (wild type) for 24 h. The results (Figure S2) indicate that these three metal-immobilized MSNs, as well as un-functionalized MSNs are non-toxic to MEF cells below a concentration of 100 μg/ml.

**Cellular uptake of MSNs**

After the study of chromobody loading and release from NTA-$M^{2+}$ complex modified MSNs, we further investigated the interactions between cells and MSNs. There are numerous studies about the cellular uptake of nanoparticles (e.g. liposomes, polyplexes and silica nanoparticles, etc.)[16, 20-27]. These studies reveal that electrostatic surface charge of nanoparticles, functionalization, particle shape and particle size are key factors that can affect the endocytosis mechanisms. For example, positively charged nanoparticles are favored for cellular uptake because the negatively charged cell membrane tends to attract the positively charged particles on its surface; particle sizes below 200 nm are considered the most favorable size for cellular internalization. In addition, the endocytosis behavior for different nanoparticles is also dependent on cell-type[28, 29]. In this study, the MSN size used for intracellular chromobody delivery is about 100 – 200 nm. The endocytosis pathways are mostly clathrin-mediated or caveolin-mediated endocytosis[16, 27]. The confocal images in Figure 3a reveal the kinetics of cellular uptake of chromobody loaded-MSNs (MSN-Ca-Cbs) by MEF-G-LMNA cells. MSN-Ca-Cbs were added to the cells in serum free medium at a concentration of 5 μg/ml, and real-time live cell images at the indicated time points were acquired. As the incubation time increased, the number of MSN-Ca-Cbs co-localized with cells also increased. After 2 h of co-incubation with MSN-Ca-Cbs, almost all cells in the imaging frame co-localized with more than two MSN-Ca-Cbs. High content statistics of cellular uptake of MSN-Ca-Cbs for different time points and concentrations were examined with a high throughput imaging system (Operetta®, PerkinElmer). Different concentrations of MSN-Ca-Cb (5 μg/ml, 10 μg/ml and 20 μg/ml) were added to each sample in a 24-well plate. At the indicated time points, cells were washed to remove free MSN-Ca-Cbs, and were then imaged by Operetta immediately. On average 600 cells were imaged and analyzed per sample. The evaluation sequence is shown in the supplementary information (Figure S3). The results (Figure 3b and 3c) from the high content analysis illustrate that cellular uptake of MSN-Ca-Cbs is time and dose dependent. After 10 min of incubation, more than 30 % of the cells have taken up MSN-Ca-Cbs already (cells with more than two MSN-Ca-Cb spots within their cytosols and nuclear region are defined as MSN-uptake cells; each MSN-Ca-Cb spot might contain more than one MSN-Ca-Cb) both for low particle concentration (5 μg/ml) and for high particle concentration (20 μg/ml). After 2 h of incubation with MSN-Ca-Cbs, more than 90 % of the cells have taken up MSN-Ca-Cbs. This result demonstrates the high cellular uptake efficiency of MSN-Ca-Cbs by MEF-G-LMNA cells.

To further visualize the MSN locations in the cytosol we used super-resolution microscopy (3D SIM). A stable MEF cell line (MEF-mEGFP) that expresses endogenous membrane co-localizing EGFP protein was developed to label the cell boundary. Cy3 labeled MSNs (MSN-Cy3) at the concentration 10 μg/ml were incubated with MEF-mEGFP cells on a glass coverslip in culture medium for 2 h. Afterwards, cells were washed to remove free MSNs and were fixed. DAPI counterstaining was performed after cell fixation. The super resolution microscope image (Figure 3d, XY view and XZ cross-section view) visually confirms that MSNs were internalized by MEF-mEGFP cells. The image clearly shows that the internalized MSN-Cy3s were equally distributed in the cytosol, and single particles can be seen in the magnified images.

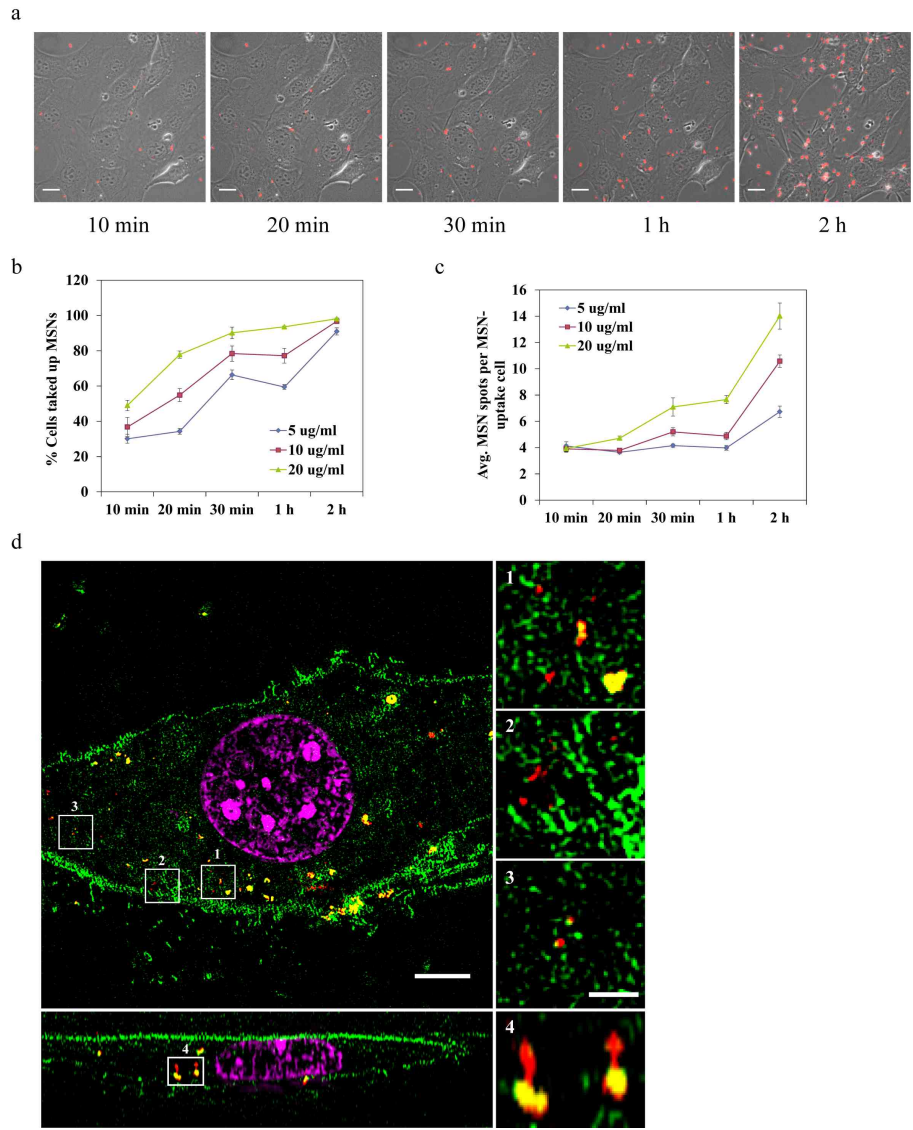

**Figure 3. Cellular uptake of MSNs.** (a) Real-time live cell imaging of the MSN-Ca-Cb uptake process by MEF-G-LMNA cells. Images were taken at the indicated time points after incubation with MSN-Ca-Cb (5 μg/ml) on MEF cells in a 2-well ibidi cell culture slide. Scar bar: 20 μm. (b), (c) Time-based analyses of cellular uptake of MSN-Ca-Cb using a high-throughput imaging system (Operetta®, PerkinElmer). Cells were seeded on 24-well plates at the concentration of $3 \times 10^4$ cells/well and incubated at 37 °C overnight. Cell culture medium was removed and different concentrations of MSN-Ca-Cb particles in serum free DMEM (phenol red free) were then added to each well. After incubation of MSN-Ca-Cb particles with cells for the indicated time intervals, free particles were removed by washing with PBS, and high-throughput live cell imaging was performed. Experiments were triplicated. An average of 600 cells were imaged in each experiment. Error bar: standard deviation. (d) Super-resolution microscopy (3D SIM) image of a cell after MSN internalization. Cy3-labeled MSNs (10 μg/ml) were incubated with MEF-mEGFP cells (MEF cells which stably express EGFP on their plasma membrane) ($3 \times 10^5$ cells/well) on a glass coverslip in a 6-well plate for 2 h. Afterwards, free MSNs were removed by washing with PBS. Cells were then fixed with 3.7 % formaldehyde and counterstained with DAPI (1 μg/ml in PBS). Green: EGFP, red: Cy3 labeled MSNs, magenta: DAPI counterstaining. Scale bar: 5 μm for main figure and XZ cross-section figure, 1 μm for magnified figures.

**Intracellular chromobody delivery**

In the intracellular chromobody delivery experiment we used MSN-Ca$^{2+}$ as the main chromobody carrier. Above we have shown that the average chromobody loading capacity in MSN-M$^{2+}$ is 70 μg/mg MSN (Figure 2c). Here we added the MSN-Ca-Cbs to the cell culture slide (μ-Slide 2-well, ibidi) at a concentration of 5 μg/ml (1 ml of culture medium per well), which yields a chromobody concentration for the intracellular delivery of approximately 25 nM. Importantly, compared to the recently published papers of direct protein delivery that use micromolar protein concentrations in each experiment, the amount of cargo we used in this work is 200 times less[5, 6]. MEF-G-LMNA cells were used to detect the release and function of GFP-specific chromobodies. When Atto 647N labeled GFP-specific chromobodies are delivered to and subsequently released from endosomes, they passively diffuse through the nuclear envelope and bind to GFP molecules on the LMNA structure. Therefore, a distinct microscopic co-localization signal from EGFP-LMNA and GFP-specific chromobodies can be observed. Figure 4 illustrates the successful delivery of GFP-specific chromobodies to MEF-G-LMNA cells. After 24 h of incubation with MSN-Ca-Cbs, several cells can be observed with chromobody release. In fact, chromobody release can already be seen after 4 h incubation of MSN-Ca-Cbs in some cells, and the chromobody staining remained on the EGFP-LMNA structure until 96 h after incubation (Figure S4). We also examined two other carriers: MSN-Ni$^{2+}$ and MSN-Zn$^{2+}$, and both of them allowed for intracellular chromobody delivery (Figure S4). To systematically quantify the efficiency of chromobody delivery via MSNs, we then used a high-throughput imaging system to image a large quantity of cells and further calculate the chromobody release efficiency (chromobody release efficiency % = chromobody stained cells/GFP-LMNA-positive cells x 100 %). The result (Figure 5a, the control group) showed only 1 – 2 % chromobody release efficiency after 24 h post MSN-Ca-Cbs incubation. We attribute the low release efficiency to endosomal trapping of most of the internalized particles, whereas the limited observed endosomal escape of chromobodies might result from the proton sponge effect generated by the His$_6$-tag on chromobodies.

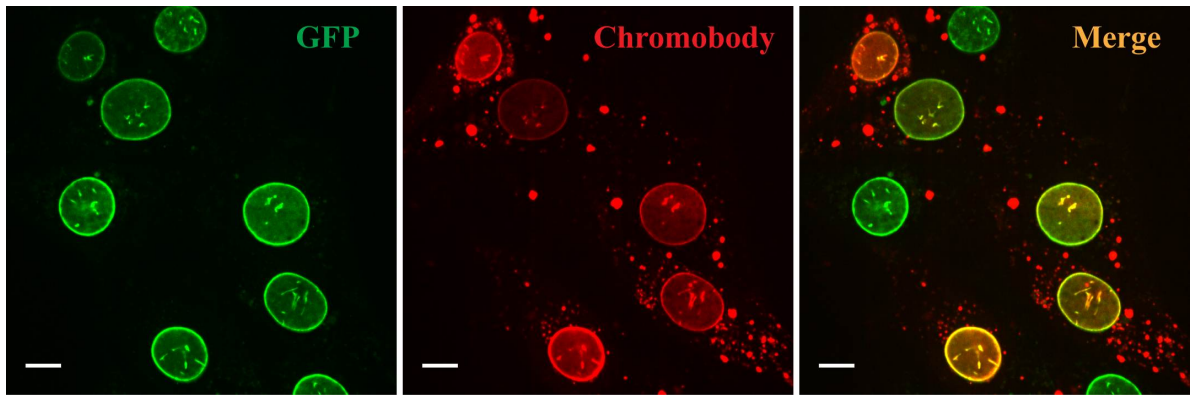

**Figure 4. Live cell confocal imaging of intracellular chromobody delivery.** 24 h post incubation of MSN-Ca-Cbs (5 μg/ml) with MEF-G-LMNA cells. Scale bar: 10 μm. Representative images are presented.

To enhance endosomal release, we investigated 4 different endosomal escape triggers: fusogenic peptide INF7, acid shock, DMSO and chloroquine.

Fusogenic peptide INF7 is a peptide derivative from influenza virus hemagglutinin HA2 protein. Wagner *et al.* first reported in 1994 that the 23 N-terminal amino acid sequence of HA2 has liposome disruption ability and erythrocyte lysis activity[30]. The derivative INF7 peptide exhibits more pH specificity with membrane disruption ability and erythrocyte lysis activity than HA2 due to its conformation change in acidic environment. Therefore, this pH-dependent conformation change and membrane-disruptive fusogenic peptide is promising for endosomal escape since the late endosomes feature an acidic environment at ~ pH 5. Here, we conjugated the INF7 peptide to MSNs using a pH-responsive acetal linker [31](Figure S5), then co-incubated MSN-INF7s with MSN-Ca-Cbs to MEF-G-LMNA cells for 2 h. However, the result (Figure 5a) showed that INF7 has no enhancement for the chromobody delivery efficiency. The second endosomal release trigger we used here is acid shock, that is, extracellular acidification of cells. The purpose of using acid shock is to generate external stress on cells and to investigate if the cellular stress response is to increase the endosomal leakage. However, the acid shock applied in this experiment had no effect on the chromobody delivery efficiency. Moreover, the acidic buffer incubation leads to 60 % cell death (Figure 5b). The third endosomal release trigger DMSO has been reported to be capable of enhancing membrane permeability[32-35]. A molecular dynamics simulation demonstrated that DMSO molecules can cause fluctuations on the two hydrophilic sides of the lipid bilayer followed by the formation of a water-permeable pore in the lipid bilayer[34]. Wang *et al*. used it to enhance the penetration efficiency of the cell-penetration peptide TAT fusion protein[35]. In our study a short incubation (5 – 10 min) with 7 % DMSO indeed enhances the release of chromobodies from the endosomes. The chromobody release efficiency increased from 1 % (control group) to 12 % after this short incubation with DMSO (Figure 5a). Furthermore, the increased endosomal release of chromobodies can be detected right after the DMSO treatment (data not shown).

The most effective endosomal release trigger studied in this work is chloroquine. Chloroquine is a well-known lysosomotropic agent (and anti-malaria drug) that preferentially accumulates in lysosomes and destabilizes the lysosomal membrane. When chloroquine is used at low concentration, it increases the pH of the acidic endosomes[36]. When applied at high concentration (> 100 μM), chloroquine can generate a strong proton sponge effect and therefore destabilize the endosomal membrane[37]. Our study shows that a short incubation of high concentration (500 μM) chloroquine facilitates endosomal release. In contrast to the DMSO treatment, the proton-sponge effect for increased endosomal release becomes obvious 24 h after the chloroquine treatment. To sum up the endosomal escape experiments, INF7 and acid shock showed no increase of chromobody release efficiency (same as the control group, 1 – 2 %) while cells treated with DMSO or chloroquine showed 12 – 18 fold increase in efficiency. The corresponding live cell images of increased chromobody release efficiency after treatment with DMSO and chloroquine are shown in Figure 5d.

Since the introduction of endosomal triggers might induce cytotoxicity, we analyzed cell viability after each endosomal trigger treatment. The cultured cells were imaged before MSN-Ca-Cbs incubation, and the numbers of GFP-LMNA-positive cells were counted. After 2 h incubation of MSN-Ca-Cbs in serum-free medium and following the treatment with endosomal release triggers, cells were imaged and counted again. The blank group refers to cells incubated in serum-free medium for 2 h without MSN-Ca-Cb treatment and no endosomal trigger treatment. Cell viability was calculated based on: GFP-LMNA-positive cells after drug treatment/GFP-LMNA-positive cells before MSN-Ca-Cb treatment x 100 %. The cell viability analysis shown in Figure 5b indicates that, compared to the blank group and the control group, the short exposure to 7 % DMSO or 500 μM chloroquine has almost no effect on cell viability. The slight decrease in cell viability of the blank group, control group, DMSO-treated cells and chloroquine-treated cells might result from the serum free medium incubation. Cell proliferation was also studied after endosomal trigger treatment to ensure the drug treatment has no effect on cell proliferation. The result in figure 5c indicated that DMSO and chloroquine treated cells proliferate normally in the following days.

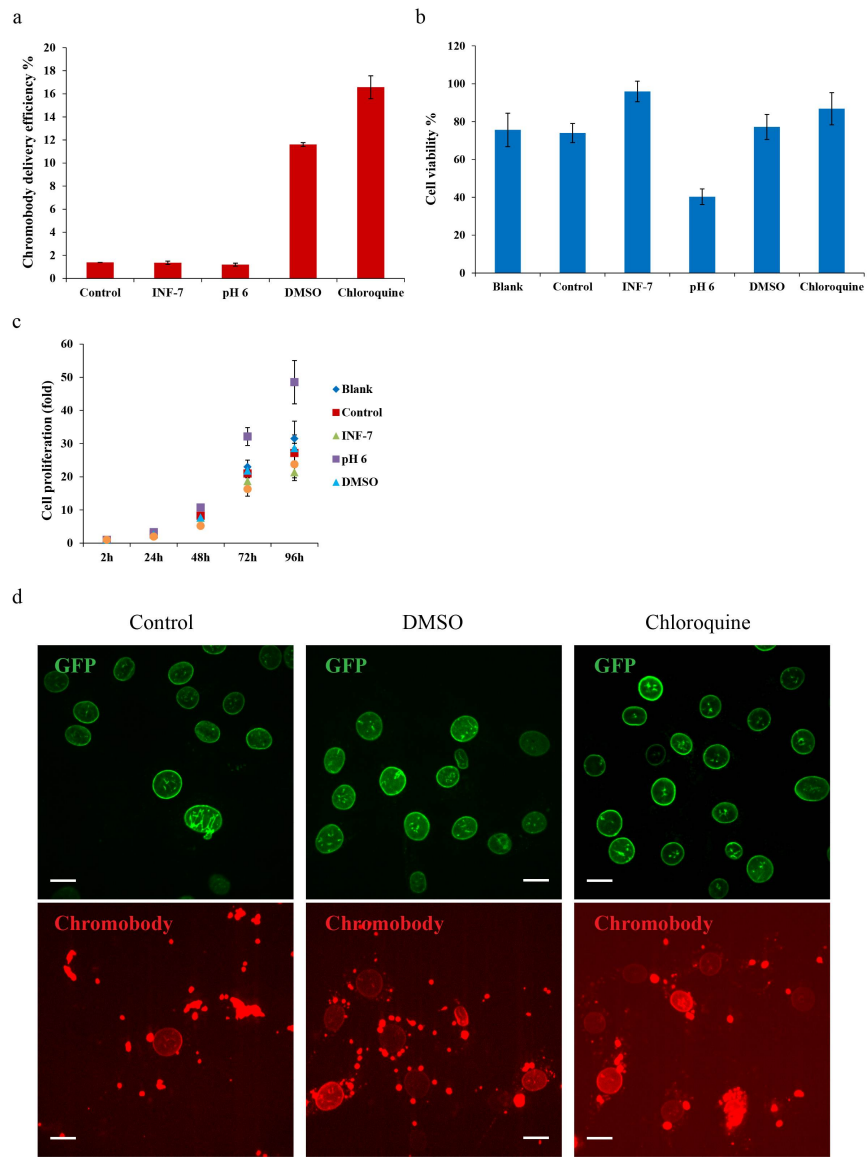

**Figure 5. Comparison of different endosomal release triggers for chromobody delivery efficiency.** High-throughput imaging system (Operetta®, PerkinElmer) was used for chromobody delivery efficiency (a), cell viability (b) and cell proliferation (c) studies. (a) Chromobody delivery efficiency after treatment with different endosomal release triggers. 24 h after MSN-Ca-Cb incubation and endosomal release trigger treatment, cells were imaged by Operetta. GFP-positive cells were counted by Harmony software, and chromobody-stained cells were counted visually. Chromobody delivery efficiency % = chromobody-stained cells/GFP-positive cells x 100 %. (b) Cell viabilities after treatment with different endosomal release triggers. Cells were imaged by Operetta before MSN-Ca-Cbs incubation. After 2 h incubation of MSN-Ca-Cbs and short treatment with different endosomal release triggers, cells in the same region of each plate were then imaged with Operetta again. Cell viability % = GFP-positive cells after drug treatment/ GFP-positive cells before drug treatment x 100 %. (c) Cell proliferations after treatment with different endosomal release triggers. The cell number immediately after drug treatment of each sample is defined as 1-fold. Cell numbers were then counted every 24 h by Operetta. (a), (b), (c) Experiments were triplicated. Error bar: standard deviation. (d) Live cell confocal images of intracellular chromobody delivery after non-treatment (control), DMSO-, and chloroquine-treated MEF-G-LMNA cells. Scale bar: 20 μm.

**Conclusion**

In this work, we successfully developed NTA-metal ion modified MSNs as carriers for intracellular chromobody delivery. We systematically studied (1) the interactions between His$_6$-tagged chromobodies and NTA-M$^{2+}$ modified MSNs; (2) cellular uptake of MSNs, and (3) endosomal release triggers for the enhancement of chromobody release efficiency. MSN-Ca$^{2+}$, MSN-Ni$^{2+}$ and MSN-Zn$^{2+}$ carriers all are applicable for successful intracellular chromobody delivery, while DMSO and chloroquine are effective triggers for drastically increased chromobody release efficiency. We could clearly demonstrate the functionality of the delivered chromobodies based on their specific binding to GFP-LMNA. Thus, the mesoporous MSN-M$^{2+}$ carriers can now enable the direct delivery of chromobodies into cells to image and understand biological processes. Moreover, we envision that the novel delivery strategy based upon the MSN-M$^{2+}$ carriers can be generalized toward the cellular delivery of other His-tagged proteins aimed at various biological applications.

**Acknowledgement**

Financial support from the Deutsche Forschungsgemeinschaft (SFB 1032 and SPP1623/LE721/13-1), Center for Nano Science (CeNS) and the Excellence Clusters Nanosystems Initiative Munich (NIM) are gratefully acknowledged. Moreover, we thank Andreas Maiser for technical assistance with the super-resolution (3D SIM) microscopy.

# References

1. Helma, J., et al., *Nanobodies and recombinant binders in cell biology.* J Cell Biol, 2015. **209**(5): p. 633-644.
2. Hamers-Casterman, C., et al., *Naturally occurring antibodies devoid of light chains.* Nature, 1993. **363**(6428): p. 446-8.
3. Rothbauer, U., et al., *Targeting and tracing antigens in live cells with fluorescent nanobodies.* Nat Methods, 2006. **3**(11): p. 887-9.
4. Helma, J., et al., *Direct and dynamic detection of HIV-1 in living cells.* PLoS One, 2012. **7**(11): p. e50026.
5. Erazo-Oliveras, A., et al., *Protein delivery into live cells by incubation with an endosomolytic agent.* Nat Methods, 2014. **11**(8): p. 861-7.
6. D'Astolfo, D.S., et al., *Efficient intracellular delivery of native proteins.* Cell, 2015. **161**(3): p. 674-90.
7. Marschall, A.L.J., et al., *Targeting antibodies to the cytoplasm.* mAbs, 2014. **3**(1): p. 3-16.
8. van Rijt, S.H., et al., *Protease-mediated release of chemotherapeutics from mesoporous silica nanoparticles to ex vivo human and mouse lung tumors.* ACS Nano, 2015. **9**(3): p. 2377-89.
9. Chen, Y.P., et al., *A new strategy for intracellular delivery of enzyme using mesoporous silica nanoparticles: superoxide dismutase.* J Am Chem Soc, 2013. **135**(4): p. 1516-23.
10. Han, D.H., et al., *Direct cellular delivery of human proteasomes to delay tau aggregation.* Nat Commun, 2014. **5**: p. 5633.
11. Meng, H., et al., *Codelivery of an optimal drug/siRNA combination using mesoporous silica nanoparticles to overcome drug resistance in breast cancer in vitro and in vivo.* ACS Nano, 2013. **7**(2): p. 994-1005.
12. Xia, T., et al., *Polyethyleneimine coating enhances the cellular uptake of mesoporous silica nanoparticles and allows safe delivery of siRNA and DNA constructs.* ACS Nano, 2009. **3**(10): p. 3273-86.
13. Slowing, II, et al., *Mesoporous silica nanoparticles as controlled release drug delivery and gene transfection carriers.* Adv Drug Deliv Rev, 2008. **60**(11): p. 1278-88.
14. Zhang, K., et al., *Facile large-scale synthesis of monodisperse mesoporous silica nanospheres with tunable pore structure.* J Am Chem Soc, 2013. **135**(7): p. 2427-30.

15. Schmidthals, K., et al., *Novel antibody derivatives for proteome and high-content analysis.* Anal Bioanal Chem, 2010. **397**(8): p. 3203-8.
16. Oh, N. and J.H. Park, *Endocytosis and exocytosis of nanoparticles in mammalian cells.* Int J Nanomedicine, 2014. **9 Suppl 1**: p. 51-63.
17. Sundberg, R.J. and R.B. Martin, *Interactions of histidine and other imidazole derivatives with transition metal ions in chemical and biological systems.* Chemical Reviews, 1974. **74**(4): p. 471-517.
18. Anderegg, G., *Critical survey of stability constants of NTA complexes.* Pure and Applied Chemistry, 1982. **54**(12).
19. Liu, Y.C., et al., *Specific and reversible immobilization of histidine-tagged proteins on functionalized silicon nanowires.* Nanotechnology, 2010. **21**(24): p. 245105.
20. Düzgüneş, N. and S. Nir, *Mechanisms and kinetics of liposome–cell interactions.* Advanced Drug Delivery Reviews, 1999. **40**(1-2): p. 3-18.
21. van der Aa, M.A., et al., *Cellular uptake of cationic polymer-DNA complexes via caveolae plays a pivotal role in gene transfection in COS-7 cells.* Pharm Res, 2007. **24**(8): p. 1590-8.
22. Rejman, J., et al., *Size-dependent internalization of particles via the pathways of clathrin- and caveolae-mediated endocytosis.* Biochem J, 2004. **377**(Pt 1): p. 159-69.
23. Chithrani, B.D. and W.C. Chan, *Elucidating the mechanism of cellular uptake and removal of protein-coated gold nanoparticles of different sizes and shapes.* Nano Lett, 2007. **7**(6): p. 1542-50.
24. Gan, Q., et al., *Effect of size on the cellular endocytosis and controlled release of mesoporous silica nanoparticles for intracellular delivery.* Biomed Microdevices, 2012. **14**(2): p. 259-70.
25. Slowing, I., B.G. Trewyn, and V.S. Lin, *Effect of surface functionalization of MCM-41-type mesoporous silica nanoparticles on the endocytosis by human cancer cells.* J Am Chem Soc, 2006. **128**(46): p. 14792-3.
26. Chung, T.H., et al., *The effect of surface charge on the uptake and biological function of mesoporous silica nanoparticles in 3T3-L1 cells and human mesenchymal stem cells.* Biomaterials, 2007. **28**(19): p. 2959-66.
27. Albanese, A., P.S. Tang, and W.C. Chan, *The effect of nanoparticle size, shape, and surface chemistry on biological systems.* Annu Rev Biomed Eng, 2012. **14**: p. 1-16.
28. Blechinger, J., et al., *Uptake kinetics and nanotoxicity of silica nanoparticles are cell type dependent.* Small, 2013. **9**(23): p. 3970-80, 3906.

29. Dobay, M.P., et al., *Cell type determines the light-induced endosomal escape kinetics of multifunctional mesoporous silica nanoparticles.* Nano Lett, 2013. **13**(3): p. 1047-52.
30. Plank, C., et al., *The influence of endosome-disruptive peptides on gene transfer using synthetic virus-like gene transfer systems.* J Biol Chem, 1994. **269**(17): p. 12918-24.
31. Schlossbauer, A., et al., *pH-responsive release of acetal-linked melittin from SBA-15 mesoporous silica.* Angew Chem Int Ed Engl, 2011. **50**(30): p. 6828-30.
32. Anchordoguy, T.J., et al., *Temperature-dependent perturbation of phospholipid bilayers by dimethylsulfoxide.* Biochimica et Biophysica Acta (BBA) - Biomembranes, 1992. **1104**(1): p. 117-122.
33. Smondyrev, A.M. and M.L. Berkowitz, *Molecular Dynamics Simulation of DPPC Bilayer in DMSO.* Biophysical Journal, 1999. **76**(5): p. 2472-2478.
34. Notman, R., et al., *Molecular basis for dimethylsulfoxide (DMSO) action on lipid membranes.* J Am Chem Soc, 2006. **128**(43): p. 13982-3.
35. Wang, H., et al., *Enhancement of TAT cell membrane penetration efficiency by dimethyl sulphoxide.* J Control Release, 2010. **143**(1): p. 64-70.
36. Mellman, I., R. Fuchs, and A. Helenius, *Acidification of the endocytic and exocytic pathways.* Annu Rev Biochem, 1986. **55**: p. 663-700.
37. Wadia, J.S., R.V. Stan, and S.F. Dowdy, *Transducible TAT-HA fusogenic peptide enhances escape of TAT-fusion proteins after lipid raft macropinocytosis.* Nat Med, 2004. **10**(3): p. 310-5.
38. Kennedy, M.J., et al., *Rapid blue-light-mediated induction of protein interactions in living cells.* Nat Methods, 2010. **7**(12): p. 973-5.

# Supporting information

# Experimental section

**Materials**

Tetraethyl orthosilicate (TEOS, Aldrich, ≥ 99 %), (3-mercaptopropyl)triethoxysilane (MPTES, Aldrich, ≥ 80 %), cetyltrimethylammonium p-toluenesulfonate (CTATos, Sigma), triethanolamine (TEA, Aldrich, 98 %), bi-distilled water is obtained from a Millipore system (Milli-Q Academic A10). 6-Maleimidohexanoic acid (Aldrich), N(alpha),N(alpha)-bis(carboxymethyl)-L-lysine hydrate (NTA-lysine, Aldrich), N-(3-dimethylaminopropyl)-N-ethylcarbodiimide (EDC, Alfa-Aesar, 98 %), N-hydroxysulfosuccinimide sodium salt (sulfo-NHS, Aldrich), iron (II) chloride tetrahydrate (Sigma Aldrich), cobalt (II) chloride hexahydrate (Aldrich), nickel chloride hexahydrate (Riedel-de Haen), copper chloride dihydrate (LMU), zinc chloride (Aldrich), calcium chloride dihydrate (LMU), magnesium chloride (Aldrich), tris(hydroxymethyl)-aminomethane (TRIS, ≥ 99 %, ROTH, 2449), acetic acid (99 % - 100 %, ROTH, 7332), thiazolyl blue tetrazolium bromide (MTT, ≥ 97.5 %, Sigma, M5655), dimethyl sulfoxide molecular biology grade (DMSO, Applichem, A3006), 3-Maleimidopropionic acid N-hydroxysuccinimide ester (99 %, Sigma Aldrich), 3,9-Bis(3-aminopropyl)-2,4,8,10-tetraoxaspiro[5.5]undecane (AK linker, Tokyo Chemical Industry), formaldehyde solution (37 %, Applichem), SSC buffer 20x (Sigma), Dulbecco's Modified Eagle's Medium (DMEM, Sigma, D6429), Dulbecco's Phosphate Buffered Saline (PBS, Sigma, D8537), FBS Superior (Biochrom, S0615), Gentamycin solution (SERVA, 50 mg/ml), trypsin-EDTA solution (Sigma, T3924), Dulbecco's Modified Eagle's Medium − phenol red free (DMEM, Sigma, D1145), L-glutamine solution (200 mM, Sigma, G7513), HEPES solution (1 M, Sigma, H3537).

**MSN synthesis and modification**

The MSN-SH was synthesized following a modified recipe reported earlier[14]. In brief, TEA (47 mg, 0.32 mmol), CTATos (0.263 g, 0.58 mmol) and $H_2O$ (13.7 g, 0.77 mmol) were mixed in a 100 ml glass flask and vigorously stirred (1250 rpm) at 80 ºC until the solution became homogeneous. A mixture of TEOS (1.8 g, 8.64 mmol) and MPTES (0.23 g, 0.96 mmol) was then added and the solution was continuously stirred (1250 rpm) at 80 ºC for 2 h. Afterwards the reaction solution was cooled down to room temperature under ambient conditions, and the particles were collected by centrifugation (43146 x g, 20 min). The organic template extraction was carried out right after the particle collection. The particle pellet was re-suspended in an ethanolic solution (100 mL) containing 2 g of ammonium nitrate, and the solution was heated to 90 °C under reflux for 1 h. The second template extraction step was subsequently performed by heating the particles under reflux at 90 °C in an ethanolic solution (100 mL) containing 10 mL of hydrochloric acid (37 %). The MSN-SH was collected by centrifugation (43146 x g, 20 min) and was washed with water and EtOH after each extraction step.

The template-extracted MSN-SH was then modified to yield NTA-functionalized MSN (MSN-NTA). 60 mg of MSN-SH and 30 mg (0.142 mmol) of 6-maleimidohexanoic acid were mixed in 12 ml of EtOH. The mixture was stirred at room temperature overnight. The resulting MSN-COOH particles were collected by centrifugation (43146 x g, 20 min), washed twice with EtOH and re-dispersed in 12 ml of EtOH. The amount of 40 mg of MSN-COOH in 8 ml of EtOH was then mixed with EDC (15 μl, 67.5 μmol) at room temperature for 10 min. Sulfo-NHS (14.6 mg, 67.5 μmol) and NTA-lysine (17.5 mg, 67.5 μmol) were mixed in 1 ml of $H_2O$ and afterwards added to the MSN solution. The mixture was stirred at room temperature for 2 h, and the resulting MSN-NTA particles were washed twice with $H_2O$ and EtOH to remove the residual chemicals. 20 mg each of MSN-SH, MSN-COOH and MSN-NTA were dried for further characterization.

**Characterization of MSN**

Transmission electron microscopy (TEM) was performed at 200 kV on a Jeol JEM-2010 instrument with a CCD detection system. A drop of diluted MSN suspension was dried on a carbon-coated copper grid at room temperature for several hours before TEM observation. Nitrogen sorption measurements were performed on a Quantachrome Instrument NOVA 4000e at 77 K. Samples (about 15 mg) were degassed at 120 °C under vacuum (10 mTorr) for 12 h before measurement. The pore volume and pore size distribution were calculated based on non-local NLDFT procedures provided by Quantachrome, using the adsorption branch of $N_2$ on silica. The hydrodynamic sizes of MSNs were measured by dynamic light scattering (DLS) analysis using a Malvern Zetasizer-Nano instrument equipped with a 4 mW He-Ne laser (633 nm). Infrared spectra of different organic functional groups on MSNs were recorded on a Thermo Scientific Nicolet iN10 IR-microscope in reflection-absorption mode with a liquid-$N_2$ cooled MCT-A detector.

**Generation of GFP-specific chromobody**

GFP-specific nanobody expression was performed in *E. coli* (JM109). Expression was induced with 0.5 mM IPTG, and cells were incubated at 37 °C for 24 h. Cells were lysed in the presence of lysozyme (100 μg/ml), DNAse (25 μg/ml) and PMSF (2 mM) followed by sonication (Branson® Sonifier; 16 x 8 sec, 20 % Amplitude) and debris centrifugation at 20000 x g for 30 min. Protein purification was performed with an Äkta FPLC system (GE Healthcare, USA) using a 5 mL His-Trap column (GE Healthcare, USA); peak fractions were concentrated to 2 ml using Amicon filter columns (cut-off 10 kDa, Merck Millipore, Germany) followed by size exclusion chromatography using a Superdex 75 column (GE Healthcare, USA). Peak fractions were pooled and protein aliquots were shock-frozen and stored at -80 °C. 1 mg purified nanobody protein was then labeled with Atto 647N (AttoTec, Germany) with a theoretical DOL (degree of labeling) of 3, according to the manufacturers' instructions. Unbound dye was removed by gel filtration in PD10 columns (GE Healthcare, USA). The final Atto 647N labelled GFP-specific chromobody was obtained and preserved in PBS in the concentration of 1 mg/ml.

**Chromobody loading and release *in vitro* test**

MSN-NTA (200 μg/vial, 6 vials) was treated with 50 mM solutions of different metal ions (0.5 mL each) ($FeCl_2$, $CoCl_2$, $NiCl_2$, $CuCl_2$, $ZnCl_2$ or $CaCl_2$) at room temperature for 6 h. After metal ion immobilization, the excess un-bound metal ions were washed out with 3 ml of $H_2O$. 200 μg of each NTA-$M^{2+}$-complex-modified MSNs (MSN-$M^{2+}$) was then suspended in 150 μl of chromobody loading buffer where the concentration of chromobody was 100 μg/ml in 0.05 M Tris-acetate buffer (pH 8.0), and incubated at 4 °C for 2 h. The un-bound chromobodies were then removed by centrifugation (4218 x g, 3 min) and the chromobody-loaded MSNs (MSN-M-Cbs) were washed 3 times (1 ml per wash) with 0.05 M Tris-acetate buffer (pH 8.0). Centrifugation (4218 x g, 3 min) was applied in every wash step to separate the supernatant and particles. Before the buffer release experiments, each type of MSN-M-Cbs was separated into two vials holding equal amounts (100 μg MSN/vial). Subsequently, 150 μl of PBS having different pH values (pH 7 and pH 5) were added to suspend the MSN-M-Cbs (100 μg). The *in vitro* chromobody release experiments were performed at 37 °C for 16 h. The solutions with released chromobody were centrifuged (4218 x g, 3 min) and then the supernatants were collected. The final chromobody release supernatants were quantified *via* their fluorescence intensity at 669 nm (excitation 644 nm) in a 96-well plate (Greiner Bio-One, Germany) by a microplate reader (Infinite® M1000 PRO, TECAN, Switzerland). The quantification was based on a calibration curve with a series of diluted pure chromobody solutions.

**Cell culture and stable cell lines**

MEF cells were cultured in DMEM medium supplemented with 10 % FBS and 50 µg/ml gentamycin under 5 % $CO_2$ at 37 °C (cell culture medium).

The plasmid construction for stable cell line generation was performed as follows. The plasmid pmEGFP-N1 was constructed by cloning of membrane EGFP [38] into pEGFP-N1 vector (Clontech) with AgeI and BsrGI endonucleases (Thermo Fisher Scientific) to substitute the EGFP with mEGFP. To prepare the pCAG-eG-LMNA-IB plasmid, the mouse LmnA/C gene was amplified with primers (forward primer: 5'-GGG CGA TCG CAT GGA GAC CCC GCT ACA and reverse primer: 5'-AGT CGC GGC CGC TTT ACA TGA TGC TGC) by PCR and cloned into plasmid under a CAG promoter with AsisI and NotI restriction sites.

To make the GFP-LMNA expression cell line, pCAG-eG-LMNA-IB was transfected into MEF cells with Lipofectamine 2000 reagent (Invitrogen), and positive cells (MEF-G-LMNA) were selected with 6 µg/ml blasticidin (Sigma) for two weeks then sorted with a fluorescence-activated cell sorting (FACS) Aria II (Becton Dickinson) instrument. For the mEGFP expression cell line, pmEGFP-N1 was transfected into MEFs, and mEGFP stable expression cells (MEF-mEGFP) were purified by FACS two weeks after transfection.

**Sample preparation for cellular uptake experiments of MSNs**

Live cell imaging medium (LCIM) containing DMEM – phenol red free, FBS (10 %), L-glutamine (200 µM), HEPES (20 mM) and gentamycin (50 µg/ml) was used for cell incubation in all live cell microscopy experiments.

For real-time tracing of cellular uptake of MSN-Ca-Cbs, MEF-G-LMNA cells were seeded on a 2-well µ-Slide (ibidi, Germany) in 50 % confluence (the proportion of the culture slide surface which is covered by cells) for overnight incubation. In each culture well, 1 ml of culture medium or LCIM was used for either cell culture or microscopy. 5 µg/ml of MSN-Ca-Cb in serum-free LCIM was added to cells and images were acquired at 10 min, 20 min, 30 min, 1 h and 2 h after the addition of MSN-Ca-Cbs.

For the high content live cell imaging, MEF-G-LMNA cells were plated on 24-well plates (Corning, USA) in 70 % confluence. The working volume for all kinds of solution (cell culture medium, PBS, etc.) in the 24-well plate is 0.5 ml per well. MSN-Ca-Cb with different concentrations (5 μg/ml, 10 μg/ml and 20 μg/ml) in serum free DMEM were added to cells. At indicated time points (10 min, 20 min, 30 min, 1 h and 2 h), cells were washed with PBS three times to remove residual MSN-Ca-Cbs in the medium. CellMask$^{TM}$ orange (Thermo Fisher Scientific) (5 mg/ml in DMSO) was diluted 1000x in LCIM and incubated with cells at 37 °C for 10 min for plasma membrane staining. After the plasma membrane staining, the CellMask solution was removed. LCIM was then added to the sample and high content imaging was performed immediately.

For the super-resolution microscopy (3D SIM), MEF-mEGFP cells were cultured on an 18 x 18 mm coverslip in a 6-well plate (Corning, USA) ($3 \times 10^5$ cells/well). 10 μg/ml of Cy3 (Lumiprobe, Germany)-labelled MSNs (MSN-Cy3) in 2 ml of serum free DMEM were added to cells and incubated for 2 h. Cells were then washed with PBS three times to remove residual MSNs, followed by fixation using 3.7 % of formaldehyde in PBS (10 min in room temperature). DAPI (1 μg/ml in PBS) counterstaining (10 min in room temperature) was performed after cell fixation. The sample was then mounted in Vectashield antifade mounting medium (Vector Laboratories, USA) to a glass slide.

**Intracellular chromobody delivery - sample preparation**

MEF-G-LMNA cells in cell culture medium were seeded on either a 2-well μ-Slide (1 ml/well) or a 24-well plate (0.5 ml/well) in 50 % confluence one day before the intracellular chromobody delivery experiment. The applied culture medium or LCIM volume in each well (1 ml/well for 2-well μ-Slide; 0.5 ml/well for 24-well plate) is the same for all following intracellular delivery experiment. MSN-Ca-Cbs, MSN-Ni-Cbs and MSN-Zn-Cbs were prepared in serum-free DMEM culture media at a concentration of 5 μg/ml, respectively. The culture medium was removed from the pre-seeded cells followed by the addition of pre-mixed MSN-M-Cb containing medium. The MSN-M-Cbs were incubated with cells at 37 °C for 2 h. Afterwards, the residual particles in the medium were washed out by PBS, and the cells were incubated in LCIM for the following live cell imaging processes.

Experiments with samples treated with endosomal triggers were carried out according to two different approaches. MSN-Ca$^{2+}$ was the main carrier used in the following experiments. For the INF7 treatment, INF7 peptide (Biosyntan GmbH) was conjugated on MSNs with pH-responsive acetal linkers as indicated in Figure S5. In brief, 10 mg (37 μmol) of 3-(Maleimido)propionic acid N-hydroxy succinimide (dissolved in 100 μl of DMSO) and 6 mg (22 μmol) of AK linker (dissolved in 1 ml of 1x SSC buffer, pH 7.4) were mixed at room temperature and stirred for 1 h. 5 mg of MSN-SH in stock solution (20 mg/ml in EtOH) was collected by centrifugation (16837 x g, 5 min) and re-suspended in previous solution. The mixture was stirred at room temperature overnight. The resulting MSN-AK-linker particles was collected by centrifugation (16837 x g, 5 min) and washed by $H_2O$ (10 ml/wash, 3 times wash). INF7 peptide consisting cysteine on its C-terminus was conjugated to 1 mg of MSN-AK-linker particle with a content of 10 INF7 molecules per MSN in 1 ml of DMSO. The mixture was stirred at room temperature for 1 h and the final MSN-INF7 particles were washed with 1 ml of DMSO two times. MSN-INF7s (1 μg/ml) and MSN-Ca-Cbs (5 μg/ml) were co-incubated with MEF-G-LMNA cells in serum free cell culture medium for 2 h. The residual particles in the incubation medium were afterwards washed out by PBS (1 ml/well). The MSN-INF7 and MSN-Ca-Cb treated cells were then incubated in LCIM for live cell imaging. For the acid shock treatment, DMSO and chloroquine release trigger tests, PBS (pH 6), 7 % DMSO in LCIM and 500 μM chloroquine in LCIM were introduced to MSN-Ca-Cb-treated cells and incubated at room temperature for 5 – 10 min. The release trigger solutions were then removed from the cells, and cells were then incubated in LCIM for live cell imaging.

**Optical microscopy**

<u>Spinning Disc</u>

MSN-M-Cb-treated cells plated on 2-well μ-Slide were selected and imaged in LCIM. 3D stacks were acquired with an UltraVIEW Vox spinning disc confocal system (PerkinElmer, UK) in a closed live cell microscopy chamber (ACU control, Olympus, Japan) heated to 37 °C, with 5 % $CO_2$ and 60 % humidified atmosphere, mounted on a Nikon Ti microscope (Nikon, Japan). Image acquisition was performed using either 63x/1.4 NA or 40x/1.3 NA Plan-Apochromat oil immersion objective lenses. Images were obtained with a cooled 14-bit EMCCD camera (C9100-50, CamLink) with a frame size of 1024 x 1024 pixels. GFP and fluorophores were excited with 488 nm (10 % power density), 561 nm (5 % power density) or 647 nm (30 % power density) solid-state diode laser lines.

<u>Operetta</u>

High-throughput images of living cells on a 24-well plate ($3 \times 10^4$ cells/well) were acquired automatically with an Operetta high content imaging system (PerkinElmer, UK).

Imaging was performed using a 40× air objective lens. GFP, CellMask and chromobody-atto647N were excited and the emissions were recorded using standard filters. The exposure time was controlled at 200 – 400 ms to avoid pixel saturation, and 50 different fields were imaged in each well. In the cellular uptake experiment, on average 600 cells were imaged. The images were then analyzed by the Harmony analysis software (PerkinElmer) with a sequence as described in Figure S3. In the endosomal release trigger, cell viability and cell proliferation analyses, MEF-G-LMNA cells before any treatment were first imaged and counted. After addition of MSNs and the following endosomal release triggers treatment, cells were imaged and counted again. The definition of cell viability % is: GFP-LMNA-positive cells after endosomal release trigger treatment/GFP-LMNA-positive cells before any treatment × 100 %. The blank group refers to the cells incubated in serum-free DMEM culture medium for 2 h without MSN and endosomal release trigger treatment. Control group refers to the cells incubated with MSN-Ca-Cbs for 2 h, but no endosomal release trigger treatment afterwards. Chromobody release efficiency was recorded 24 h after MSN-Ca-Cbs introduction. Cells exhibiting a chromobody release signal were counted visually. The chromobody release efficiency is defined as: chromobody-LMNA-positive cells/GFP-LMNA-positive cells × 100 %. Cells with different treatments were imaged and counted every 24 h for the cell proliferation calculation. 400 to 10000 cells were imaged in the endosomal release trigger, cell viability and cell proliferation tests depending on the recording time point.

Super-resolution microscopy (3D-SIM)

Super-resolution microscopy (3D-SIM) was performed with a Delta Vision OMX v3 (Applied Precision, GE Healthcare) instrument equipped with 405 nm, 488 nm and 593 nm laser diodes, a 100×/1.4 NA Plan-Apochromat oil immersion objective lens (Olympus) and Cascade II:512 EMCCD cameras (Photometrics). The image stacks were acquired with a z-step of 125 nm and with 15 images per plane (five phases, three angles). The raw data were computationally reconstructed by using the SoftWorx 5.1.0 software.

**MTT assay**

MEF cells (wild type) were first seeded on 96-well microplates with the concentration of $5 \times 10^3$ cells/well in cell culture medium (100 μl/well), and incubated at 37 °C for 16 h. MSNs (un-MSNs, MSN-Ni$^{2+}$, MSN-Zn$^{2+}$, MSN-Ca$^{2+}$) (2 mg from stock solution) in EtOH were centrifuged down (16873 x g, 5 min), and re-suspended in cell culture medium at the concentration of 1 mg/ml. The MSN solutions were then series diluted to the desired particle concentrations. After removal of the culture medium, 100 μl of MSN solution with different concentrations was added to each well. Also, 100 μl of cell culture medium was added to cells to serve as control group. The wells with 100 μl of cell culture medium but without cells were referred to as blank groups. MSN solutions and cells were then co-incubated at 37 °C overnight. Afterwards, the MSN-treated cells were washed with PBS buffer three times to remove the residual particles. MTT was diluted in cell culture medium at the concentration of 0.5 mg/ml and then added to the cells (100 μl/well). After incubating the MTT solution and cells for 4 h, purple crystals metabolized by healthy cells were observed. Subsequently, 100 μl of pure DMSO was added to each well and the samples were incubated at 37 °C for 1 h until the purple crystals were dissolved. The absorbance at 570 nm and the reference absorbance at 655 nm were measured on each sample using a Microplate reader (Infinite® M1000 PRO, TECAN).

# Supplementary figures

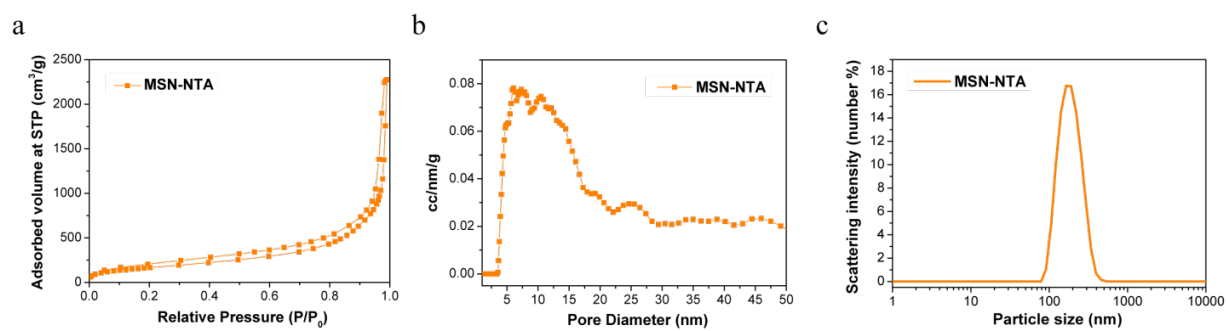

**Figure S1. Characterization of NTA-conjugated MSNs (MSN-NTA).** (a) Nitrogen sorption isotherm, (b) pore size distribution and (c) DLS analysis (particles suspended in EtOH) indicated that MSN-NTA preserved the large-pore mesoporosity and colloidal stability after surface modification.

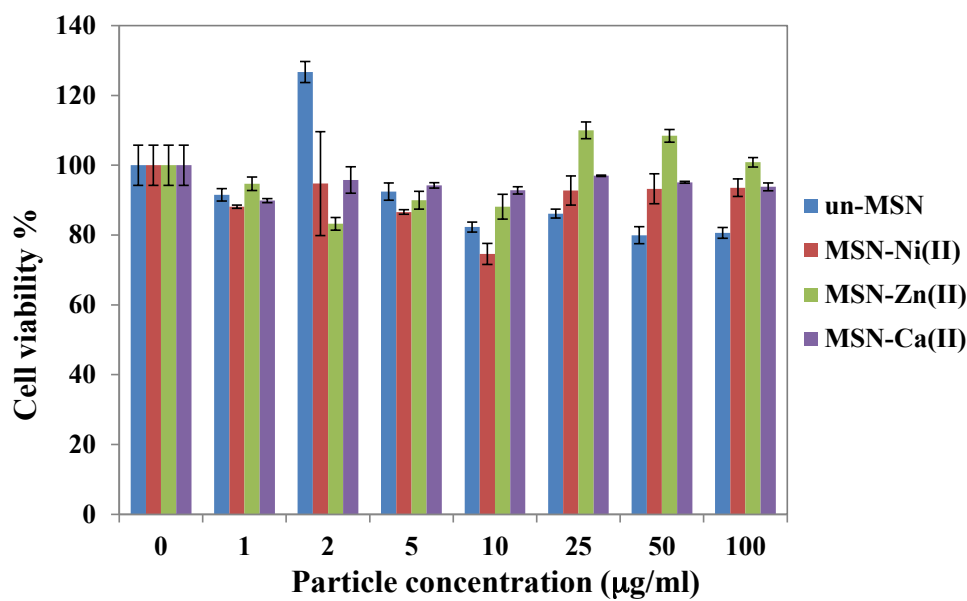

**Figure S2. Cytotoxicity tests (MTT assay) of different metal-treated MSN-NTA samples (MSN-Ni$^{2+}$, MSN-Zn$^{2+}$ and MSN-Ca$^{2+}$) and un-functionalized MSN (un-MSN) on MEF cells (wild type).** All the particle types show no significant cytotoxicity below the particle concentration of 100 μg/ml.

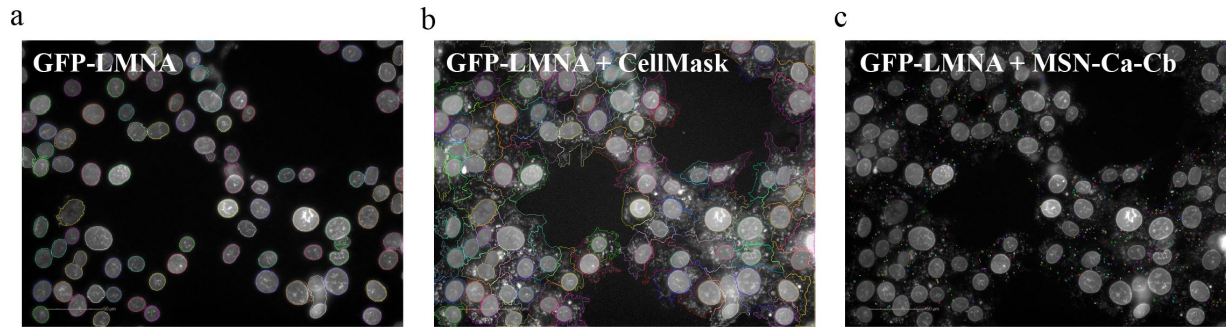

**Figure S3. High throughput imaging evaluation of cellular uptake of MSN-Ca-Cbs.** MEF-G-LMNA living cells were stained with CellMask orange and imaged at indicated time points after the addition of MSN-Ca-Cbs. Fluorescence from GFP, CellMask orange, and Atto 647N (labeled on chromobodies) were recorded separately with Operetta high content image analysis system using standard filter for 488 nm, 546 nm and 647 nm emission. Images were analyzed with Harmony analysis software as the following sequence. (a) The nuclear region is segmented from the background according to GFP-LMNA signal where the GFP-LMNA signal is shown in grey and the circular color lines stand for the segmentation results. Most of the recognized nuclear region fit the GFP-LMNA signal, which indicates a correct segmentation. (b) The cytoplasm region was recognized *via* CellMask signal segmentation. The cytoplasm region is obtained by subtraction of nuclear region obtained from step (a). The border of the segmented cell region was shown as closed color lines around the nuclear region. (c) Recognition of MSN-Ca-Cbs taken up by the cells. Internalized MSN-Ca-Cbs were visualized as spots within the cytoplasm region, and the spots recognition was presented as color dots in the image. The segmented results (nuclei, cytoplasm and MSN-Ca-Cb spots) were used to define populations. Population 1: Cell = GFP-LMNA nucleus plus its surrounding cytoplasm region. Population 2: MSN-uptake cell = cell with more than two MSN-Ca-Cb spots in its cytoplasm and nucleus region. The evaluation results were based on the calculation: (i) % cell take up MSN-Ca-Cbs = MSN-uptake cells/total cells; (ii) average MSN-Ca-Cb spots per MSN-uptake cell = total spots number in all MSN-uptake cells/MSN-uptake cells.

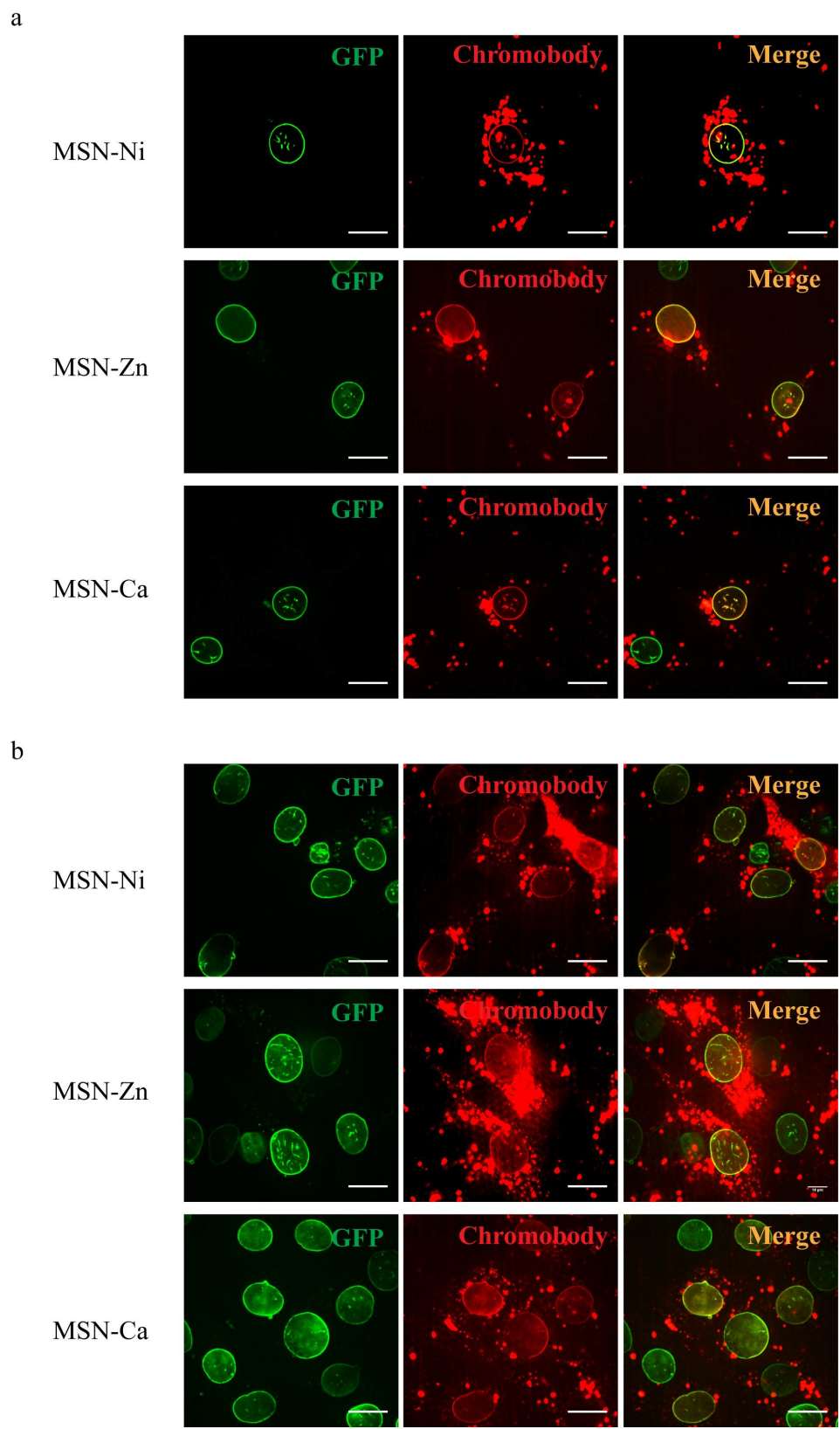

**Figure S4. Intracellular delivery of chromobodies *via* different metal-treated MSNs.** (a) 4 h after MSN-Ni-Cb/MSN-Zn-Cb/MSN-Ca-Cb incubation. (b) 96 h after MSN-Ni-Cb/MSN-Zn-Cb/MSN-Ca-Cb incubation.

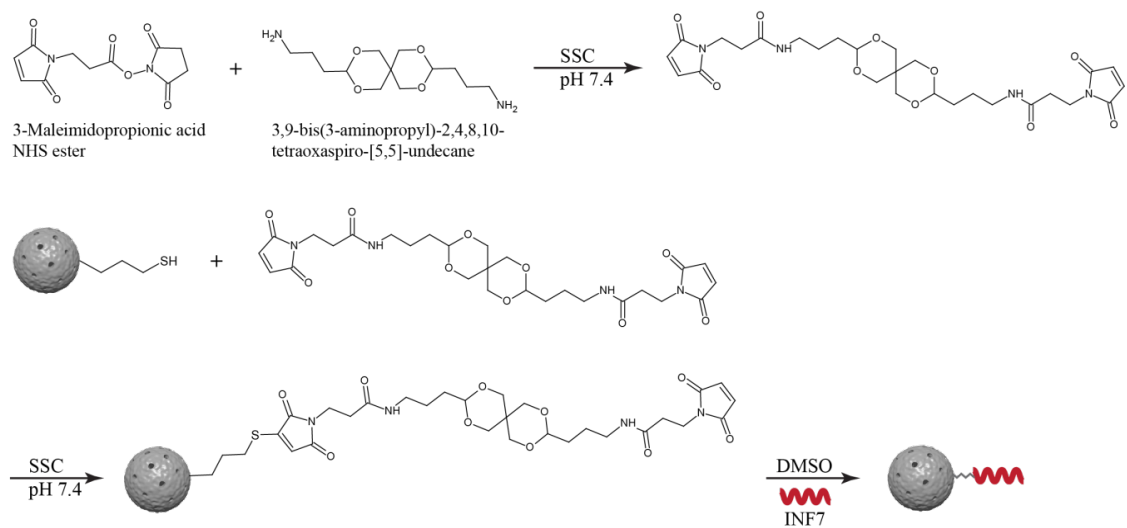

**Figure S5. Conjugation of INF7 peptide to MSNs *via* pH-responsive acetal linker.** MSN-SH was conjugated with pH-responsive acetal linker (3,9-bis(3-aminopropyl)-2,4,8,10-tetraoxaspiro-[5,5]-undecane) (AK linker) *via* the maleimide-NHS heterobifunctional crosslinker (3-Maleimidopropionic acid NHS ester) yielding MSN-AK-linker. INF7 peptide consisting of cysteine on its C-terminus was covalently attached to MSN-AK-linker through the maleimide-thiol reaction.